\def\cm{{\rm\thinspace cm}}

\def\erg{{\rm\thinspace erg}}

\def\keV{{\rm\thinspace keV}}
\def\km{{\rm\thinspace km}}

\def\Mpc{{\rm\thinspace Mpc}}
\def\Msun{\hbox{$\rm\thinspace M_{\odot}$}}

\def\s{{\rm\thinspace s}}

\def\kmpspmp{\hbox{$\km\s^{-1}\Mpc^{-1}\,$}}

\def\pcmsq{\hbox{$\cm^{-2}\,$}}

\documentclass[usegraphicx]{mn2e}

\usepackage{fixltx2e}
\usepackage{mathptmx}

\include{defn}
\begin{document}

\title[Long XMM observation of IRAS13224-3809]{Long XMM observation of
  the Narrow-Line Seyfert 1 galaxy IRAS13224-3809: rapid variability,
  high spin and a soft lag } \author[A.C. Fabian et al]
{\parbox[]{6.5in}{{ A.C. Fabian$^1\thanks{E-mail: acf@ast.cam.ac.uk}$,
      E.~Kara$^1$, D.~Walton$^1$, D. Wilkins$^1$, R.R.~Ross$^{12}$,
      K.~Lozanov$^1$, P.~Uttley$^2$, L.~Gallo$^3$, A.~Zoghbi$^4$,
      G.~Miniutti$^5$, T.~Boller$^6$, W.N.~Brandt$^7$,
      E.M.~Cackett$^8$, C-Y.~Chiang$^1$, T.~Dwelly$^9$, J.~Malzac$^1$
      J.M.~Miller$^{10}$, E.~Nardini$^{11}$, G.~Ponti$^9$,
      R.C. Reis$^{10}$, C.S.~Reynolds$^4$, J.~Steiner$^1$,
      Y.~Tanaka$^6$ and
      A.J.~Young$^{12}$ }\\
    \footnotesize $^1$ Institute of Astronomy, Madingley Road,
    Cambridge
    CB3 0HA\\
    $^3$ Department of Astronomy and Physics, Saint
    Mary’s University, Halifax, NS B3H 3C3, Canada\\
    $^4$ Dept. of Astronomy, University of Maryland,
    College Park, MD 20742, USA\\
    $^5$ Centro de Astrobiologia (CSIC-INTA), Dep. de Astrofisica;
    LAEFF, PO Box 78, E-28691, Villanueva de la Ca{\~n}ada, Madrid,
    Spain\\
    $^6$ Max-Planck-Institute for Extraterrestrial Physics, Garching, PSF 1312, 85741 Garching, Germany\\
    $^7$ Department of Astronomy and Astrophysics, Pennsylvania State
    University, 525 Davey Lab, University Park,
    PA 16802, USA\\
    $^8$ Dept. of Physics and Astronomy,
    Wayne State University, Detroit, MI 48201, USA\\
    $^9$ School of Physics and Astronomy, University of Southampton,
    Highfield, Southampton SO17 1BJ\\
    $^{10}$ Dept. of Astronomy, University of Michigan, Ann Arbor,
    MI 48109, USA\\
    $^{11}$ Harvard-Smithsonian Center for Astrophysics, 60 Garden
    Street, Cambridge, MA 02138, USA\\
    $^{12}$ H. H. Wills Physics Laboratory, University of Bristol, Tyndall Avenue, Bristol BS8 1TL, UK\\
  }}

\maketitle
  
\begin{abstract}
  Results are presented from a 500~ks long XMM-Newton observation of
  the Narrow-Line Seyfert 1 galaxy IRAS\,13224-3809. The source is
  rapidly variable on timescales down to a few 100~s. The spectrum
  shows strong broad Fe-K and L emission features which are
  interpreted as arising from reflection from the inner parts of an
  accretion disc around a rapidly spinning black hole. Assuming a
  power-law emissivity for the reflected flux and that the innermost
  radius corresponds to the innermost stable circular orbit, the black
  hole spin is measured to be 0.988 with a statistical precision
  better than one per cent. Systematic uncertainties are discussed. A
  soft X-ray lag of 100\,s confirms this scenario. The bulk of the
  power-law continuum source is located at a radius of 2--3
  gravitational radii.
 
\end{abstract}

\begin{keywords}
 accretion, accretion discs -- black hole physics -- line:profiles -- 
X-rays:general 
\end{keywords}

\section{Introduction}

The X-ray emission from most Narrow-Line Seyfert 1 galaxies (NLS1) is
characterised by a steep soft X-ray spectrum and rapid
variability. The most extreme such objects are 1H\,0707-495 and
IRAS13228-3809, which both show a sharp drop above 7~keV in XMM Newton
spectra (Boller et al 2002, 2003). 1H\,0707-495 has been further
studied several times with XMM, including a long 500~ks dataset in
2008 which revealed broad iron K and L lines and a soft lag of about
30~s (Fabian et al 2009; Zoghbi et al 2010, 2011). In contrast,
IRAS13224-3809 only had 64~ks of XMM data (Ponti et al 2010; Gallo et
al 2004), despite showing spectacular variability during ROSAT (Boller
et al 1997) and ASCA (Dewangan et al 2002) observations. New
observations totalling 500~ks have now been made with XMM in 2011 and
reported here.

The unusual spectrum and 7~keV drop of both objects have been
interpreted as due to either intervening absorption or strong
relativistic blurring of a reflection component (Boller et al 2002,
2003; Fabian et al 2004).  1H\,0707-495 dropped into a low state for
about 2 months at the start of 2011 during which an XMM spectrum
showed evidence for even more blurring. The results are consistent
with the power-law component of the X-ray source lying within one
gravitational radius of the central black hole (Fabian et al 2012). In
the normal state, one third of this component extends to $\sim
20r_{\rm g}$. 

The combination of the above results with the reverberation lags in
1H\,0707-495 and in over a dozen other sources (Emanoulopolous,
McHardy \& Papdakis 2010; Tripathi et al 2011; De Marco et al 2011,
2012; Zoghbi \& Fabian 2011 and Zoghbi et al 2012) provides very strong
support for the reflection model for the X-ray emission of Seyfert
galaxies. In this model the primary power-law component lies above the
inner accretion disc around the black hole and produces the X-ray
reflection component by irradiation of the disc (see e.g. Fabian \& Ross
2010). The soft lags are then the light travel time difference between
the power-law and reflection components as detected by the
observer. The new data presented in this paper are interpreted within
the reflection model.

IRAS13224-3809 is a radio quiet (1.4~GHz flux of 5.4~mJy, Feain et al
2009) NLS1 at redshift $z=0.066$. For a flat $\Lambda$CDM cosmology
with $H_0=71\kmpspmp$, its luminosity distance is 293~Mpc.
\section{Observations and Data Reduction}

IRAS~13224-3809 was observed for $\sim 500$~ks with the {\em
  XMM-Newton} satellite (Jansen et al 2001) from 2011 July 19 to 2011
July 29 (Obs. IDs 0673580101, 0673580201, 0673580301, 0673580401).  We
focus on the data from the EPIC-pn camera (Str\"uder et al 2001). The
first observation was taken in full window imaging mode, and the
following three in large window imaging mode.  All of the data were
reduced in the same way, using the {\em XMM-Newton} Science Analysis
System (SAS v.11.0.0) and the newest calibration files.

The data were cleaned for high background flares, resulting in a final
total exposure of 300~ks.  The data were selected using the condition
{\sc pattern} $\le 4$. Pile-up effects were not significant in any of
the observations.

\begin{figure*}
  \centering
  \includegraphics[width=0.8\textwidth,angle=0]{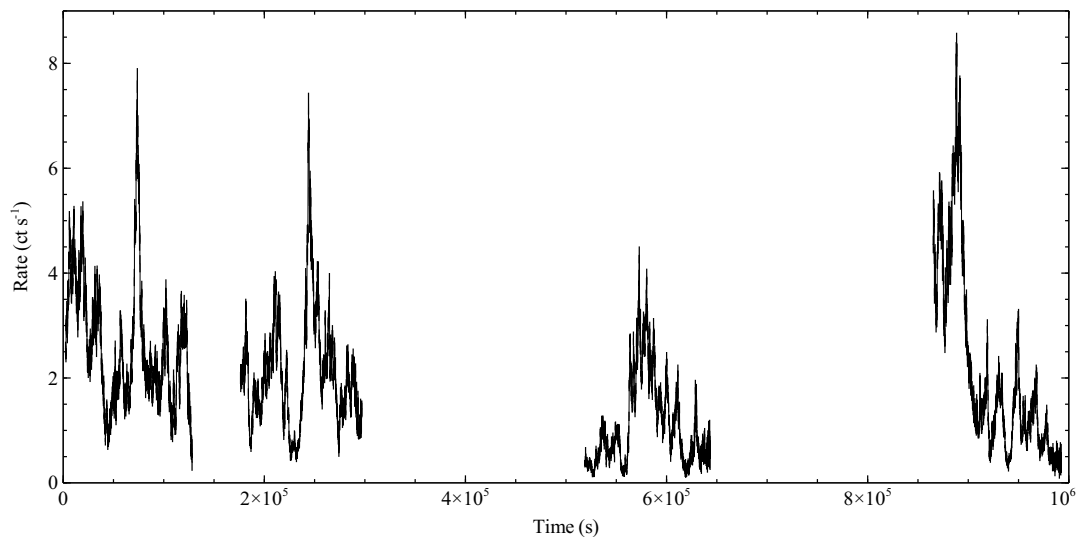}
  \caption{Full band XMM lightcurve (0.3--10~keV) lightcurve of
    IRAS13224-3809. Bins are 200~s. }
\end{figure*}

\begin{figure}
  \centering
  \includegraphics[width=0.9\columnwidth,angle=0]{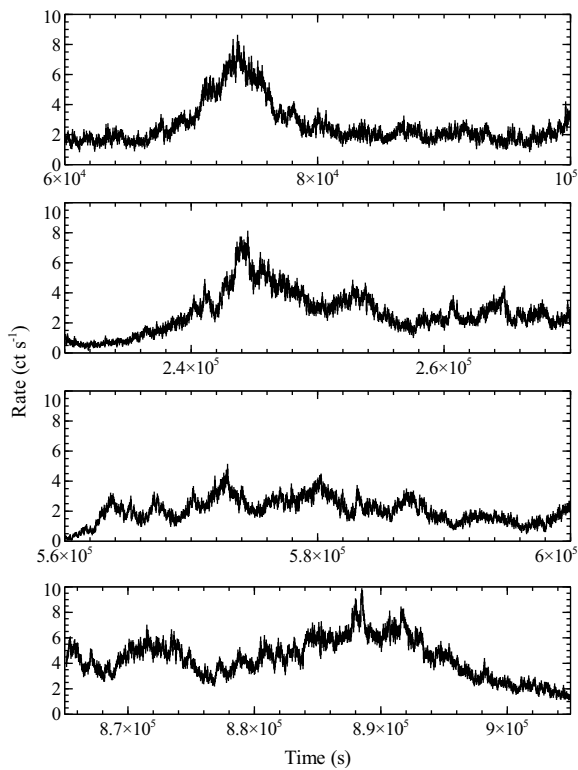}
  \caption{Lightcurves of regions of peak count rate per orbit in the
    0.3--10~keV band. Bins are 50~s.  }
\end{figure}

\begin{figure*}
  \centering
  \includegraphics[width=0.9\textwidth,angle=0]{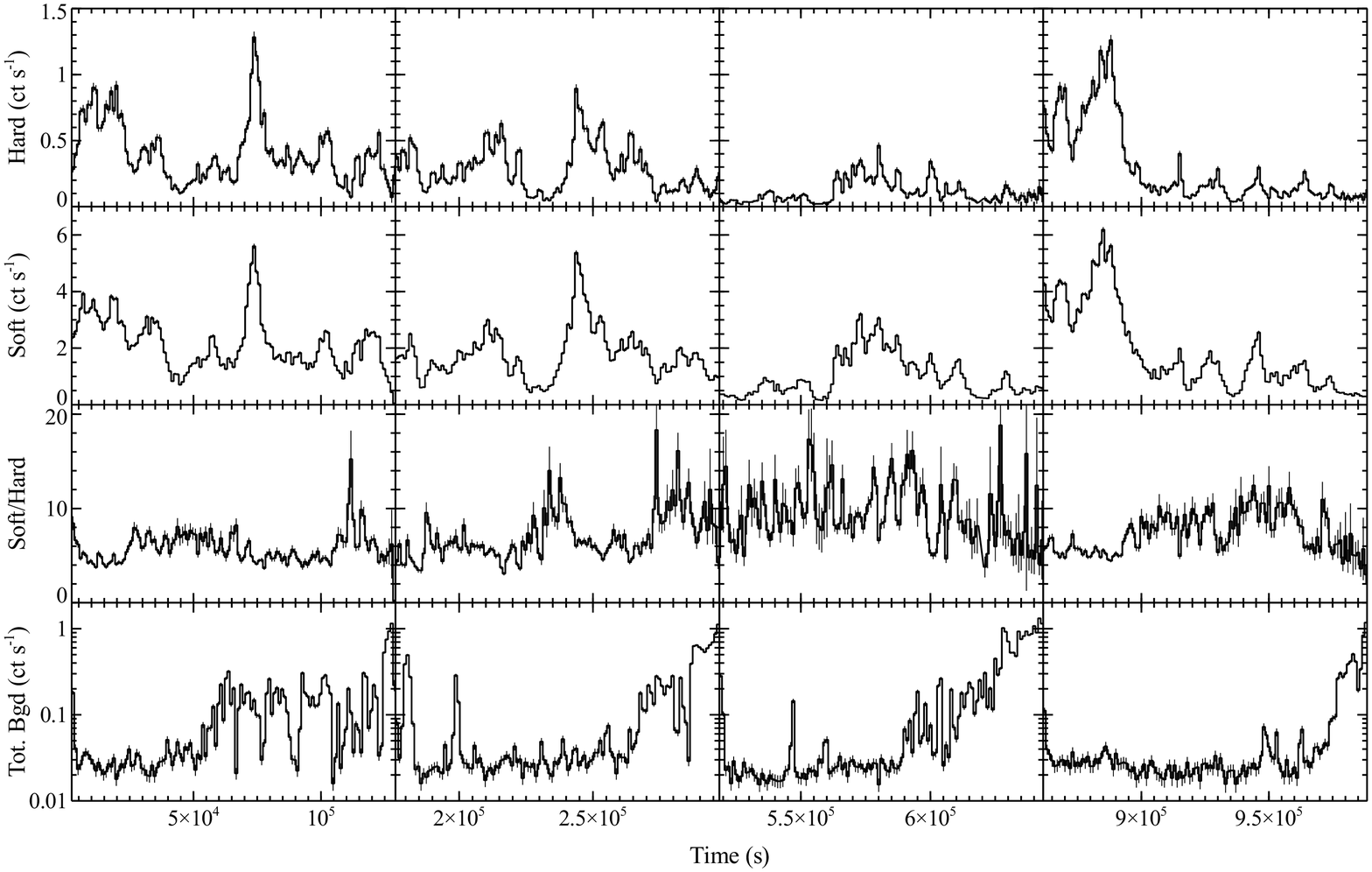}
  \caption{Hard (1--4~keV) and soft band (0.3--1~keV) light curves
  (top two panels), softness ratio and log (total background
  rate) (lower two panels). Bins are 1000~s.}
\end{figure*}

The source spectra were extracted from circular regions of radius 35
arcsec, which were centered on the maximum source emission, and the
background spectra were chosen from a circular region of the same size
and on the same chip. The position of the background regions were
chosen to avoid the Cu-K emission lines from the electronic circuits
behind the pn CCD that contaminate the background at 8.0 and
8.9~keV.  The response matrices were produced using {\sc rmfgen} and
{\sc arfgen} in {\sc SAS}.

The spectra from the four observations were merged before fitting
using {\sc mathpha} in {\sc FTOOLS}, and the resulting combined
spectrum was rebinned to contain a minimum of 20 counts per bin.
Spectral fitting was performed using {\sc xspec} v12.5.0
(Arnaud 1996). Quoted errors correspond to 90\% confidence
level. Energies are given in the rest frame of the source. Quoted
abundances refer to the solar abundances in Anders \& Grevesse (1989).

\section{Lightcurve and Variability}

The lightcurve of the long XMM observation in 2011 is shown in
Fig.~1. Start and stop dates are 2011 July 19 and 29. The source is
clearly highly variable with several pronounced upward spikes of
emission. The observation consists of 4 orbits of XMM, the first two
of which are contiguous. Gaskell (2003) used ASCA data to show that
the X-ray light curves can be lognormal, in the sense that a frequency
histogram of log(count rate) is gaussian. We find that this is only a
fair description of Orbit 1 and is a poor description of the Orbits 3
and 4.

\begin{figure}
  \centering
  \includegraphics[width=0.9\columnwidth,angle=0]{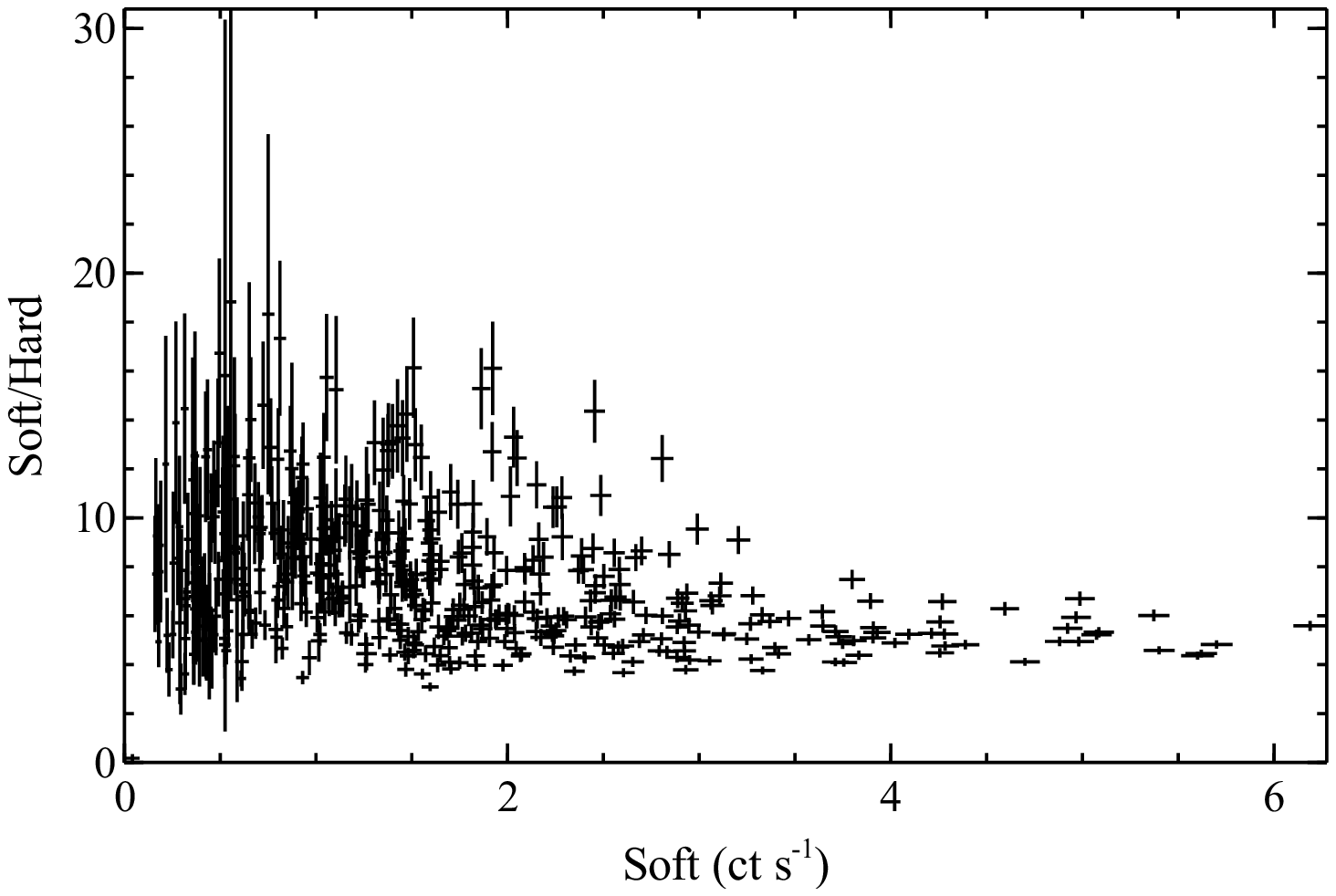}
  \caption{Softness ratio plotted versus Soft count rate.  }
\end{figure}

The light curve is reminiscent of that seen with the ROSAT HRI, reported
by Boller et al (1997). Observations were made every day for a month
resulting in a total exposure of $\sim110$~ks and mean observation
length $\sim3$~ks. 5 flares of emission were found rising above about
150~ct~ks$^{-1}$. Using {\sc webpimms} (Mukai 1993) and a 0.1(0.2)~keV
blackbody model (see next Section for detailed spectrum) we find that
150~ct~ks$^{-1}$ corresponds to about 6(3.4) EPIC~pn~ct~s$^{-1}$. We
see 3 flares above 6~pn~ct~$s^{-1}$ and $\sim11$ above 3.4~pn~ct~s$^{-1}$ in
500~ks. The numbers  depend on the precise spectral shape,
but are similar enough to indicate that the behaviour of the source is
similar to that in 1997.

The bright spikes of emission are shown at higher time resolution in
Fig.~2. The most rapid large rise occurs in Orbit 2, where the count
rate jumps by by about 4.5~ct~s$^{-1}$ from $(2.42-2.44)\times
10^5$~s, i.e. 2000~s. Using the spectral model developed in Section 3,
this corresponds to about $7\times 10^{40}\erg$~s$^{-2}$ in the
0.3--2~keV band. An even faster event occurs at the peak of Orbit 4
where the count rate jumps up and down by 3~ct~s$^{-1}$ in just a few
100~s. This means that active regions of the source at that time must
be smaller than a few 100 light seconds in size.  The rate of change
of luminosity is thus approaching the value of $10^{42}\erg$~s$^{-2}$,
especially if the energy band is extended down to 0.1~keV, which is
the highest previously recorded, by Brandt et al (1999). That paper
discusses the lower limit on the radiative efficiency required by the
source in converting mass to energy to produce such a luminosity
gradient (Fabian 1979); the efficiency must be high and approaching 50
per cent. This is difficult to envisage taking place without invoking
non-spherical geometry and/or relativistic effects.

Light curves in both Soft (0.3--1~keV) and Hard (1--4~keV) bands are
shown in Fig.~3, together with the Soft/Hard ratio and background
count rate (note that the log of the background rate is shown as the
values are mostly small).  The pronounced large spikes in count rate
show no strong spectral variation. A trend for the spectrum to be
sometimes softer when the source is faint is apparent (Fig.~4). Orbit 3
shows considerable hardness ratio variations that do not seem to
correlate with the count rate of either the source or background.

\section {Spectral fits}

Following the phenomeonological approach of Fabian et al (2009; see
also Ponti et al 2010), we fit the spectrum over the 0.3--0.4~keV and
1.2--2.2~keV bands with a simple absorbed power-law plus blackbody
model and then show the residuals to that model over the full
0.3--9~keV band (Fig.~4). Clear, broad emission residuals
corresponding to the iron K and L bands are apparent.  A good fit can
be made over this range with that simple model plus two
relativistically blurred lines, at 0.92 and 6.7~keV using the {\sc
  Laor} model. The blurring parameters are tied between the two lines,
yielding an inclination of $\sim60$~deg and an emissivity index of
6.6.  The equivalent width of the lines are 59.6~eV and
1.74~keV. Untying the parameters leads to a slightly better fit but unphysical
energies (1.02 and 7.06~keV) and a lower inclination ($\sim30$~deg).

In practice we do not expect that the emission peaks are due to single
lines but to line and absorption complexes in the reflection
spectrum. We have therefore fitted the data with a physical model consisting
of blackbody, power-law and two reflection components, one of high and
the other of low ionization, similar to the best fitting model for
1H\,0707-495. The motivation for the two ionization components is to
model better a turbulent accretion disc.  The model used is {\tt
  phabs*zphabs*(blackbody+
  kdblur*(atable\{reflionx.mod\}+atable\{reflionx.mod\})}, where the
relativistic-blurring convolution model {\sc kdblur} acts on the
ionized reflection model {\sc reflionx} of Ross \& Fabian 2005)). The
results of this fit are shown in Table~2 and Figs.~6 and 7.

\begin{figure}
  \centering
  \includegraphics[width=0.65\columnwidth,angle=-90]{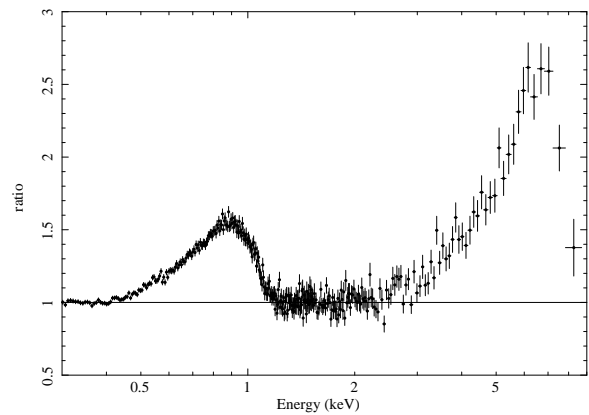}
  \caption{Ratio of observed spectrum to a model spectrum. The model
    consists of a power-law, blackbody and two Laor broad lines which
    have been are fitted to the data. The normalizations of the Laor
    lines have been set to zero before displaying. }
\end{figure}

\begin{figure}
  \centering
  \includegraphics[width=0.57\columnwidth,angle=-90]{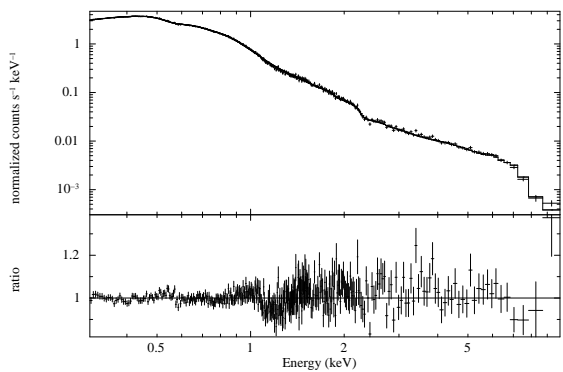}
  \caption{Full band pn spectrum fitted with double reflectors and a
    blackbody component.  }
\end{figure}

\begin{figure}
  \centering
  \includegraphics[width=0.6\columnwidth,angle=-90]{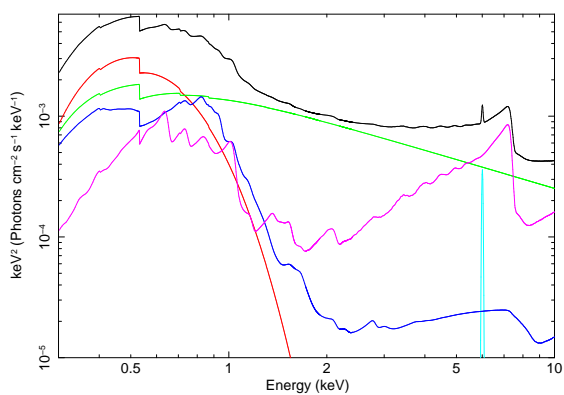}
  \caption{Components of the best-fitting model shown in Fig.~6.  }
\end{figure}

\begin{table}
  \caption{Values of  model parameters used in emisivity profile
    determination. The absorption component
    is  fixed at the Galactic value. }
\centering
\begin{tabular}{llll}
  	\hline
   	\textbf{Component} & \textbf{Parameter} & \textbf{Value}
         \\
	\hline
        Absorption & Galactic $N_{\rm H}\pcmsq$ & $5.3 \times 10^{20}$  \\
	powerlaw  & Photon index, $\Gamma$ & $2.70^{+0.007}_{-0.01}$ 
        \\
        & Norm &  $3.86\times 10^{-4}$  \\
	\hline
	{\sc relconv}  & Inclination, $i$\,deg & $63.8\pm0.4$  \\
	& $R_{\rm br}$\,r$_{\rm g}$ & $2.1\pm0.3$  \\
        & Inner Index, $q_1$ & $>9$  \\
        & Outer Index, $q_2$ & $3.4^{+0.05}_{0.1}$ \\
        & Spin, $a$ & $0.988\pm 0.001$\\
        \hline
        blackbody & temperature, $kT$\,keV & $0.103\pm0.0007$
          \\
        & Norm & $3.65\times 10^{-5}$ \\
        \hline
	{\sc extendx}  	& Iron abundance / solar & $>16$  \\
	& Ionization parameter, $\xi_1$ & $20.7\pm0.4$  \\
        & Ionization parameter, $\xi_2$ & $325^{+38}_{-11}$  \\
        & Norm$_1$ & $3.0\times 10^{-8}$ \\ 
        & Norm$_2$ & $5.0\times 10^{-6}$ \\ 
	& Redshift, $z$ & $4.06\times 10^{-2}$ \\
	\hline
        &$\chi^2/{\rm dof}$ & $955/906$  \\
        \hline
\end{tabular}

\label{par.tab}

\end{table}

We obtain the best spectral fit if the higher ionization component is
replaced by a gaussian line at 0.86~keV (which is then
relativistically blurred along with the lowly-ionized component). The
need fro any absorption edge now disappears. This suggests that the
spectral model we are using is incomplete. It is possible  that the
blackbody component is part of the disc emission and helps heat the
disc, so increasing the collisional part of the Fe-L emission which,
for a temperature of 0.3~keV (the peak of the blackbody) peaks between
0.8 and 0.9~keV. Generating the appropriate grids of models for this is
beyond the scope of the present work.

\subsection{The spin of the black hole}

The spectral fit shown in Fig.~6 requires a steep emissivity profile
from $\sim 1.35 r_{\rm g}$ which, if identified as the innermost
stable circular orbit (ISCO), means that the black hole is close to
maximal spin.  Fitting the spectrum with the blurring kernel {\sc
  reflconv} (Dauser et al 2010) we find the spin to be $0.988\pm0.001$
(see also Table~1).
\begin{figure}
  \centering
  \includegraphics[width=0.7\columnwidth,angle=-90]{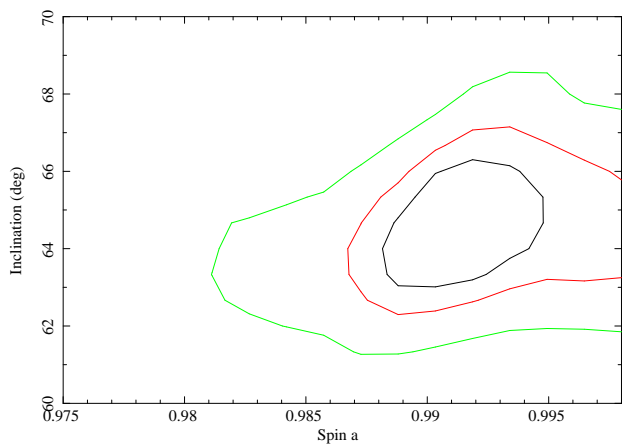}
  \caption{68, 90 and 99 per cent confidence contours for spin and
    disc inclination.}
\end{figure}

The value of spin is robust to changes in the model with differences
in $a$ being less than one per cent in all models tried. A small
improvement in $\chi^2$ is obtained by including a cold iron line
(rest energy 6.4~keV); a slightly broad one with $\sigma=0.23\keV$
being preferred. Including a small edge at 1.15~keV produces a more
significant drop in $\chi^2$. This may indicate the presence of a hot
wind, as found for 1H\,0707-495 by Dauser et al (2012), or just be a
deficiency in the model since it occurs where the reflection
components are dropping steeply (particularly see Fig.~5). The
variation of spin with inclination is shown in Fig.~8, where no strong
correlation between these parameters is evident.

Although the statistical uncertainty on the measured spin is well
below one per cent, there are larger systematic uncertainties which
have yet to be determined. The most important is perhaps the implicit
identification of the innermost radius of the reflector with the
ISCO. Computations suggest that the uncertainty here is small and
could be less than $0.5r_{\rm g}$ (Reynolds \& Fabian 2008; Shafee et
al 2008). The work of Schnittmann et al (2012) emphasises that
emission from matter on plunge orbits is beamed mostly into the black
hole. We note that the requirement for a low ionization component
emphasises that the disc remains dense and thus thin within the final
gravitational radius.

\subsection{Inferring the position and size of the power-law source}

The break in the emissivity profile at only $\sim 2.1 r_{\rm g}$
indicates that the power-law source is close to the black hole, within
a few gravitational radii, and thus must be small and confined within
that radius (e.g. Wilkins \& Fabian 2011, 2012).

Confirmation that the source is very close to the black hole comes
from the reflection fraction. This is the ratio of the reflection
components to the power-law component, normalized so that unity
corresponds to a reflector subtending $2\pi$~sr. This is not
straightforward to calculate for a high $\Gamma$ source since the {\sc
  reflionx} model does not tabulate the total flux, but only that
above 0.1~keV (the flux at lower energies is of course included in the
computations). We assess the reflection fraction by comparing the
ratio of the amplitude of the Compton hump around 30~keV of the low
ionization reflection component to the power law with that predicted
by the {\sc pexrav} model. The result is a reflection fraction of
about 15, which is a strong indication of light bending close to the
black hole (Martocchia \& Matt 1996; Miniutti \& Fabian 2004).

The emissivity profile has been determined in more detail by fitting
the spectrum above 3~keV by the sum of relativistically-blurred
emission profiles from contiguous radii of the disc (see Wilkins \&
Fabian 2010 for more details; the energy range is restricted to the
Fe-K band as the Fe-L band consists of many overlapping emission
lines). The result (Fig.~9, top panel) indicates a triple power-law
emissivity profile, which integrates to show where the observed
photons originate from on the disc (Fig.~9, lower panel). 80 per cent
of the photons are reflected within about $2.5 r_{\rm g}$ with the
remaining 20 per cent mostly coming from within $10-20 r_{\rm g}$.
\begin{figure}
  \centering
  \includegraphics[width=\columnwidth,angle=0]{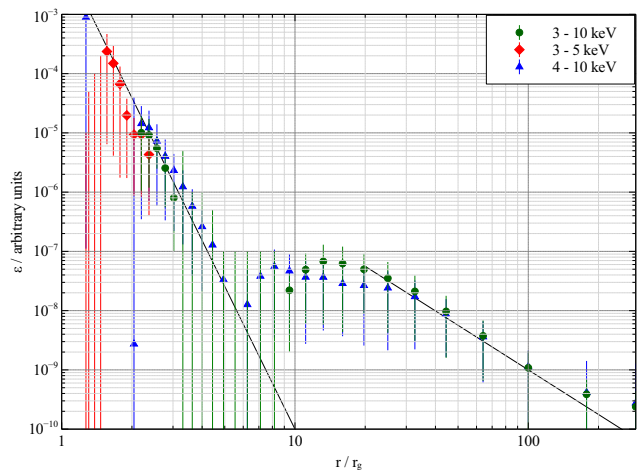}
  \includegraphics[width=\columnwidth,angle=0]{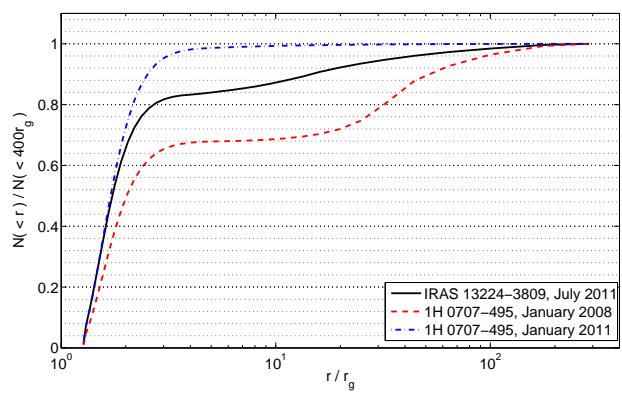}
  \caption{Top: Emissivity profile obtained by fitting the data in the
  broadened Fe-K band with emission from many annuli. Bottom:
  Integrated photon flux as function of radius.}
\end{figure}

Motivated by this, in order to estimate the size and location of the
major primary X-ray source, we generated a grid of emissivity profiles
for a range of cylindrical X-ray source regions of varying radial
extent and at varying heights above the plane of the accretion disc
using the high speed GPU-based general relativistic ray tracing code
of Wilkins \& Fabian 2012. The grid consists of sources extending
between 1 and $50 r_{\rm g}$ radially, the bases of which are between
1.0 and $3.1 r_{\rm g}$ above the plane of the accretion
disc. Initially, the thickness of the source is set to be $0.5 r_{\rm
  g}$.

These emissivity profiles were then fitted to the profile of the Fe-K
emission line using a modified version of the {\sc KDBLUR} convolution
model, leading to the constraints shown in Fig. 10. These fits imply
the source is either radially extended to $\sim1 r_{\rm g}$ at a
height of $h\sim 2-25 r_{\rm g}$ or extended out to around $2-3.5
r_{\rm g}$ at a height of $\sim 1.7 r_{\rm g}$.

\begin{figure}
  \centering
  \includegraphics[width=0.6\columnwidth,angle=-90]{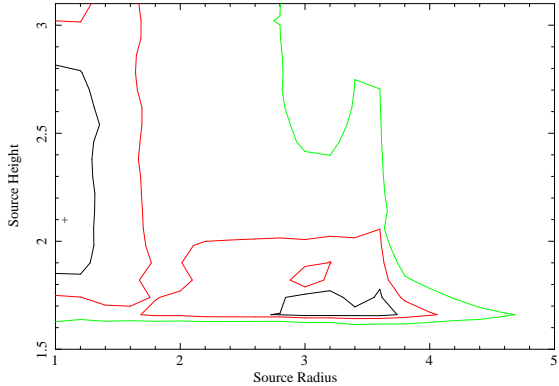}
  \caption{68, 90 and 99 per cent confidence contours for the radius
    and height (lower edge) of a slab source of thickness $0.5 {\rm r_g}$.  }
\end{figure}

The integrated emissivity profile of IRAS\,13224-3809 is compared with
the normal and low state profiles of 1H\,0707-495 in Fig.~9. It
appears to be sandwiched between the two. This could be either due to
the source always having two primary emission components, one compact
as shown in Fig.~10 the second a more extended component, or to the
source varying in size with time, the larger source likely being
associated with the brighter phases. This will be investigated in
later work. 

\section{Rapid variability and the soft lag}

The fractional RMS variability spectrum, computed according to the
prescription of Edelson et al (2002), is shown in Fig.~11. It
resembles that of many other sources in which reflection is present,
resembling a combination of variable power-law and reflection
components. The amplitude of the variability of the power-law
component needs to be greater than that of the reflection in order that
the broad Fe-K line appears inerted in this Figure.

Using the light curves of the four orbits, we compute the Fourier
phase lag between the hard and soft energy bands, following the
technique described in (Nowak 1999).  The background-subtracted
light curve segments range in length from $8.34 \times 10^{5}$~s to
$1.24 \times 10^{5}$~s with 10~s bins.  The soft band is defined from
0.3 -- 1 keV, where the soft-excess dominates the spectrum. The hard
band, 1.2 -- 5 keV, is dominated by emission from the power law
continuum.  From the Fourier transforms of the hard and soft band
light curves, $\widetilde{S}$ and $\widetilde{H}$ respectively, we
compute their phase difference, $\phi(f) = \mathrm{arg}[\langle
\widetilde{H}^{\ast}\widetilde{S} \rangle]$, where $\ast$ denotes
complex conjugate. We convert this to a frequency-dependent time lag,
$\tau(f) \equiv \phi(f)/2\pi f$.  Using this sign convention, a
negative lag means that the soft band light curve lags behind the hard
band.

The results are show in the lag-frequency spectrum in
Fig.~\ref{lag_freq}.  The hard flux lags behind the soft by hundreds
of seconds at frequencies less than $\sim~2 \times 10^{-4}$~Hz.
At frequencies $\nu
\sim~[3 - 5] \times 100\s$. The light-crossing time of $2 r_{\rm g}$
for a mass of $5\times 10^6\Msun$ is $\sim 50\s$, so a total lag of
100~s or so is reasonable.

\begin{figure}
  \centering
  \includegraphics[width=\columnwidth,angle=0]{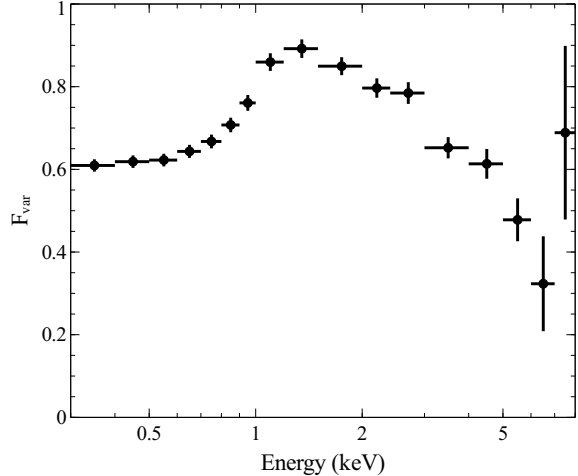}
  \caption{Fractional RMS variability spectrum using 500~s bins. . }
\end{figure}

\begin{figure}
\includegraphics[width=\columnwidth]{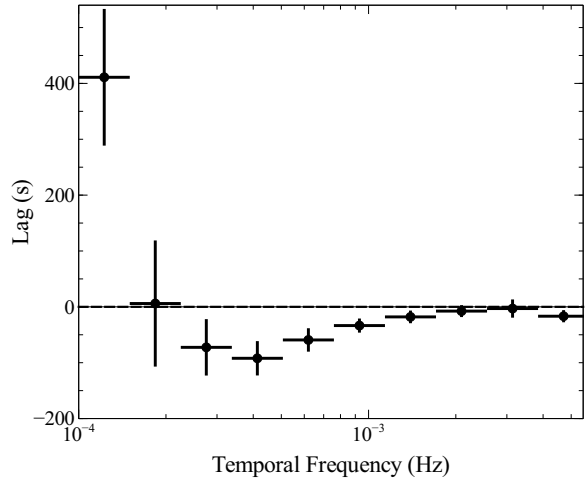}
\caption{Lag-frequency spectrum for this 500~ks observation. The lag
  is calculated between the soft energy band (0.3 -- 1.~keV) and the
  hard band (1.2 -- 4.~keV). We adopt the convention that negative lag
  mean the soft band lags behind the hard band. The most negative lag
  (at $3.4 \times 10^{-4}$~Hz) is $-92.1 \pm 30.7$~s. }
\label{lag_freq}
\end{figure}

\section{Discussion}

IRAS\,13224-3809 is remarkably similar in overall X-ray behaviour to
1H\,0707-495. The variability of IRAS\,13224-3809 may be the most
extreme. We shall explore the behaviour of the source as a function of
time and flux in more detail in later work.  

The X-ray spectra of both sources require high iron abundance ($A_{\rm
  Fe}\sim 10-20$).  In recent work, Wang et al. (2012) have presented
a strong correlation between metallicity, as measured by the
Si~\textsc{iv} O~\textsc{iv}~]~/~C~\textsc{iv} ratio, and outflow
strength in quasars, as obtained via the blueshift and asymmetry index
(BAI) of the C~\textsc{iv} emission line. Their results indicate
highly significant super--solar metallicity ($Z/Z_\odot \geq 5$) for
quasars with BAI$\geq 0.7$. This results indicates that metallicity
likely plays an important role in the formation and acceleration of
quasar outflows as expected, for instance, if quasar outflows are
predominantly line--driven.

\begin{figure}
\begin{center}
\includegraphics[width=0.35\textwidth,height=0.45\textwidth,angle=-90]{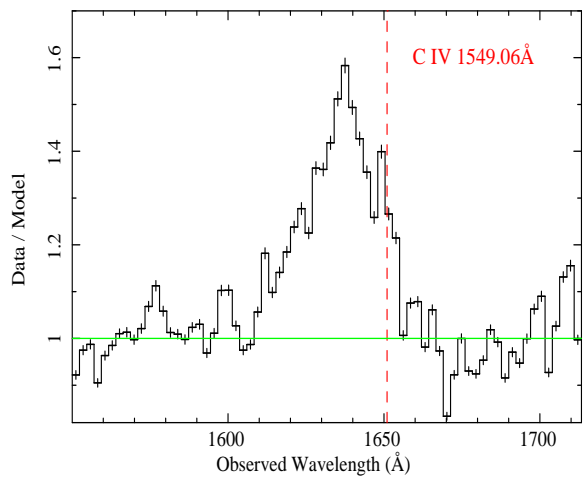}
\caption{The C~\textsc{iv} emission line profile from the HST--STIS
  observation performed on June 1999 with the G~140L grating is shown
  in the observed frame. Data have been slightly rebinned for visual
  clarity. The vertical line shows the expected wavelength of the
  C~\textsc{iv} emission line for a redshift $z=0.0658$.}
\label{BAI}
\end{center}
\end{figure}

As mentioned above, both IRAS~13224--3809 and 1H~0707--495 are
characterised by extremely blueshifted C~\textsc{iv} emission lines
with almost no contribution at rest wavelength. Their UV spectra
indicate that BAI$\geq 0.9$ in both sources, as shown in
Fig.~\ref{BAI}. If the metallicity--BAI correlation of Wang et
al. (2012) extends or saturates above their largest observed BAI
($\sim 0.76$), one infers that IRAS~13224--3809 and 1H~0707--495 are
characterised by $Z/Z_\odot \geq 8$. A strong indication for
super--solar metallicity in both sources is consistent with the strong
FeII lines in the optical spectra and was also inferred by Leighly (2004)
via photoionisation modelling of the UV spectra.

A $\sim 100\s$ soft lag is detected, which is a direct prediction of
the reflection modelling used for the source. With the many other lags
now seen, this justifies the reflection spectrum approach. It is
consistent with the spectral modelling which indicates that the bulk
of the primary continuum emission source is only a few gravitational
radii in size and distance from the black hole. The spin of the black
hole is high and close to maximal. This may be the result of secular
evolution dominating in Narrow-Line Seyfert 1 galaxies, as inferred by
Orban de Xivry et al (2011).

\section*{Acknowledgements}
 ACF thanks
the Royal Society for support.  RCR thanks the Michigan Society of
Fellows and NASA for support through the Einstein Fellowship Program,
grant number PF1-120087. EK is supported by the Gates Cambridge Scholarship.


\begin{thebibliography}{}
\bibitem[]{} Anders E., Grevesse N., 1989, GeCoA, 53, 197
\bibitem[]{} Arnaud K.A., 1996, ASPC, 101, 17
\bibitem[]{} Boller T., Brandt W.N., Fabian A.C., Fink H.H., 1997,
  MNRAS, 289, 393 
\bibitem[]{} Boller T., Fabian A.C., Sunyaev R., Tr\"umper J., Vaughan
  S., Ballantyne D.R., Brandt W.N., Keil R., Iwasawa K., 2002, 329, 1
\bibitem[]{} Boller T., Tanaka Y., Fabian A., Brandt W.N., Gallo L.,
  Anabuki N., Haba Y., Vaughan S., 2003, MNRAS, 343, L89
\bibitem[]{} Brandt W.N., Boller T., Fabian A.C., Ruszkowski M., 1999,
  MNRAS, 303, L53
\bibitem[]{} Dauser T., Wilms J., Reynolds C.S., Brenneman L.W., 2010,
  MNRAS, 409, 1534
\bibitem[]{} De Marco B., Ponti G., Uttley P., Cappi M., Dadina M.,
  Fabian A.C., Miniutti G., 2011, MNRAS submitted
\bibitem[]{} Dewangan G.C., Boller T., Singh K.P., Leighly K.M., 2002,
  A\&A, 390, 65
\bibitem[]{} Edelson R., Turner T.J., Pounds K., Vaughan S., Markowitz
  A., Marshall H., Dobbie P., Warwick R., 2002, ApJ, 568, 610
\bibitem[]{} Emmanoulopoulos D., McHardy I.M., Papadakis I.E., 2011,
  MNRAS, 416, L94
\bibitem[]{} Fabian A.C., 1979, Proc. Roy. Soc., 366, 449
\bibitem[]{} Fabian A.C., Miniutti G., Gallo L., Boller T., Tanaka Y.,
  Vaughan S., Ross R.R., 2004, MNRAS, 353, 1071
\bibitem[]{} Fabian A.C., et al 2009, Nature, 459, 540
\bibitem[]{} Fabian A.C., Ross R.R., 2010, Sp. Sci. Rev., 157, 167
\bibitem[]{} Fabian A.C. et al 2012, MNRAS, 419, 116 
\bibitem[]{} Feain I.J. et al 2009, ApJ, 707, 114
\bibitem[]{} Gallo L., Boller T., Tanaka Y., Fabian A.C., Brandt W.N.,
  Welsh W.F., Anabuki N., Haba Y., 2004, MNRAS, 347, 269
\bibitem[]{} Gaskell M., 2004, ApJ, 612, L21
\bibitem[]{}Jansen F., et al., 2001, A\&A, 365, L1
\bibitem[Leighly \& Moore(2004)]{2004ApJ...611..107L} Leighly, K.~M.,
  Moore, J.~R.\ 2004, ApJ, 611, 107
\bibitem[Leighly(2004)]{2004ApJ...611..125L} Leighly, K.~M.\ 2004,
  ApJ, 611, 125
\bibitem[]{} Martocchia A., Matt G., 1996, MNRAS, 282, L53
\bibitem[]{} Miniutti G., Fabian A.C., 2004, MNRAS, 349, 1435
\bibitem[]{} Mukai, K. 1993, Legacy 3, 21-31
\bibitem[]{} Nowak M.~A. Vaughan B., Wilms, J., Dove J.B., Begeman
  M.C., 1999, ApJ, 510, 874
\bibitem[]{} Orban de Xivry G., Davies R., Scartmann M., Komossa S.,
  Marconi A., Hicks E., Engel H., Tacconi L., 2011, MNRAS, 417, 2721
\bibitem[]{} Ponti G. et al 2010, MNRAS, 406, 2591
\bibitem[]{} Reynolds C.S., Fabian A.C., 2008, ApJ, 675, 1048
\bibitem[]{} Ross R.R., Fabian A.C., 2005, MNRAS
\bibitem[]{} Shafee R., Narayan R., McClintock J., 2008, ApJ, 676, 549
\bibitem[]{} Str{\"u}der L., et al., 2001, A\&A, 365, L18
\bibitem[]{} Tripathi S., Misra R., Dewangan G., Rastogi S., 2011, ApJL
 736, L37
\bibitem[Wang et al.(2012)]{2012ApJ...751L..23W} Wang, H., Zhou, H., Yuan,
W., Wang, T.\ 2012, ApJ, 751, L23
\bibitem[]{} Wilkins D.R., Fabian A.C., 2010, MNRAS, 414, 1269 
\bibitem[]{} Wilkins D.R., Fabian A.C., 2011, MNRAS, 424, 1284 
\bibitem[]{} Zoghbi A., Fabian A.C., Uttley P., Miniutti G., Gallo
  L.C., Reynolds C.S., Miller J.M., Ponti G., 2010, MNRAS, 401, 2419
\bibitem[]{} Zoghbi A., Uttley P., Fabian A.C., 2011, MNRAS, 412, 59
\bibitem[]{} Zoghbi A., Fabian A.C., 2011, MNRAS. 418, 2642
\bibitem[]{} Zoghbi A., Fabian A.~C., Reynolds C.~S., Cackett E.~M.,
  2012, MNRAS, 422, 129
\end{thebibliography}
\end{document}